\documentclass[pre,twocolumn,superscriptaddress,showpacs]{revtex4}
\usepackage{amssymb}
\usepackage{graphicx}
\usepackage{subfigure}
\begin{document}

\title{Fractal dimension and threshold properties in a spatially 
correlated percolation model}
\author{Hongting Yang}
\affiliation{Department of Physics and Astronomy, University of Southern
California, Los Angeles, CA 90089-0484}
 
\affiliation{Department of Physical Science and Technology, Wuhan
University of Technology, Wuhan  430070, P.R. China}
\author{Wen Zhang}
\affiliation{Department of Physics and Astronomy, University of Southern
California, Los Angeles, CA 90089-0484}

\author{Noah Bray-Ali} 
\affiliation{Department of Physics and Astronomy, University of Southern
California, Los Angeles, CA 90089-0484}
\affiliation{Department of Physics and Astronomy, University of Kentucky, Lexington, KY 40506}

\author{Stephan Haas}
\affiliation{Department of Physics and Astronomy, University of Southern
California, Los Angeles, CA 90089-0484}
\date{\today}

\begin{abstract}
We consider the effects of spatial correlations in a two-dimensional site 
percolation model. By generalizing the Newman-Ziff Monte Carlo algorithm to include spatial correlations, 
percolation thresholds and fractal dimensions of percolation clusters are obtained.  For a wide range of spatial correlations, the percolation threshold differs little from the uncorrelated result.  In contrast, the fractal dimension differs sharply from the 
uncorrelated result for almost all types of correlation studied.  We interpret these results in the framework of long-range correlated percolation. 
\end{abstract}

\maketitle

\section{Introduction}
Percolation refers to the formation of long-range connectivity across a lattice or network, and applies to various complex physical systems, 
including
aerogel\cite{MG1995}, polymers\cite{BH1995}, 
superconductors\cite{FS2003,S2004,DM2005,A2006}, 
and magnets\cite{YR2005,HV2006}. Outside physics, percolation also arises in the context of 
ecology\cite{WC1995}, complex networks\cite{NW1999,MN2000,CE2000,CN2000}  
and social science\cite{SW2000,SC2004,S2005}.

In these applications, sites or bonds are assumed to randomly occupy an underlying lattice or network.  At a critical concentration, the percolation threshold $p_c$, a connected cluster of sites spans the system.  If the occupation probabilities on different sites are independent, the problem maps onto the $q$-state Potts model in the limit $q\rightarrow 1$\cite{SA1992}.  In two dimensions, the critical properties of the $q$-state Potts 
model can be obtained from an equivalent Coulomb gas system for arbitrary $q$\cite{dutch}.  In the limit $q\rightarrow 1$, one obtains the percolation length exponent $\nu=\frac{4}{3}$ and the fractal dimension $D_f=\frac{91}{48}$ of the percolation cluster appearing at threshold.  A rigorous derivation of the same results follows from a map of percolation onto Schramm-Loewner evolution with $\kappa=6$\cite{SLE}. 

In physical applications, it is common for there to be correlations between the occupation probabilities on different sites.  This case can also be mapped to a $q$-state Potts model\cite{W1984}; however, the resulting model has correlated coupling-constant disorder.  For short-range correlations, the disorder is irrelevant and does not change universal properties such as $\nu$ and $D_f$.  Interestingly, this is also true in the case where the correlations decay asymptotically as $x^{-a}$ for sufficiently large distances $x$, provided that the exponent satisfies $a>D$ where $D$ is the number of 
spatial dimensions\cite{W1984}.

For long-range correlations, the critical properties may change.  In particular, for power-law correlations satisfying the extended Harris criterion $a\nu-2<0$, a long-range correlated renormalization-group fixed point has been predicted to occur with new 
percolation length exponent $\nu_{long}=\frac{2}{a}$ and new fractal dimension \cite{W1984}.  Although estimates have been made of the fractal dimension near the upper critical dimension $D=6$, no general predictions or numerical work exist in two dimensions, the lower critical dimension.  This work provides one of the first estimates of the fractal dimension in the presence of long-range correlations in two dimensions.  Prior work has evaluated the fractal dimension of critical Ising clusters on the triangular lattice, a particular case of long-range correlated site percolation with $a=\frac{1}{8}$\cite{W1984}.

We focus on spatial correlations in a site 
percolation model on a square lattice. 
There are various ways of implementing 
correlations. For example, in an earlier study\cite{DR1974,D1978}, a 
vacant site was assigned a probability $F^yR$ for being occupied. Here $F$ 
is a number representing enhancement,  $R$ is the probability for 
occupation for each of those vacant sites not having an occupied neighbor, 
and $y$ is the number of its occupied neighbors. $F=1$ corresponds to the 
uncorrelated percolation situation; for $F>1$, the vacant sites with more 
occupied neighbors have larger occupation probabilities than the others. 

This model can be used to simulate correlated absorption. An interesting 
result found for this model was that the percolation threshold $p_c$ is a 
non-monotonic function of the correlation strength $F$. In particular, 
correlations between sites do not always promote the occurrence of 
percolation, but in some cases instead shift the percolation threshold 
to larger occupation densities compared to the uncorrelated case.

To investigate long-range correlations, a more efficient algorithm for growing the percolation clusters is needed.  A few years ago, Newman and Ziff developed a fast Monte Carlo algorithm 
for site and bond percolation\cite{NZ2001}. In this algorithm, a sample 
state with $n+1$ occupied sites or bonds is realized by adding one extra 
randomly chosen site or bond to a sample state with $n$ sites or bonds. In 
this way, a relatively small amount of computational effort is needed to 
determine the new clusters from the old. Here, we generalize the  
Newman-Ziff algorithm to the case of correlated percolation.

\section{Growth process}
For simplicity, let us define the distance between nearest neighbor sites 
to be one unit and introduce an adjustable integer length parameter $d$. 
In each configuration, starting from an empty square lattice, only the 
first occupied site is completely randomly chosen. The second site to be 
occupied is determined by first randomly choosing a vacant site, and then 
checking the projections of its distance to the first occupied site both 
along the x-axis and y-axis. 

If both projections are no larger than $d$, 
we occupy the vacant site. Otherwise, we randomly choose another vacant 
site and check its distance from the initial site, until finally both  
projections are smaller than or equal to $d$. In a similar way we occupy 
the other vacant sites. The correlated distribution is thus obtained by 
ensuring that there exists at least one occupied site with its distance to 
the next occupied site no larger than $d$, both in the horizontal and the 
vertical direction. A main difference of this correlated site percolation 
model to earlier ones\cite{DR1974,D1978} is that we do not permit the 
occupation of any vacant sites farther than $d$ away from the previously 
occupied sites.

\begin{figure}[h]
\begin{center}
\subfigure[Correlated]{\includegraphics[width=0.4\textwidth]{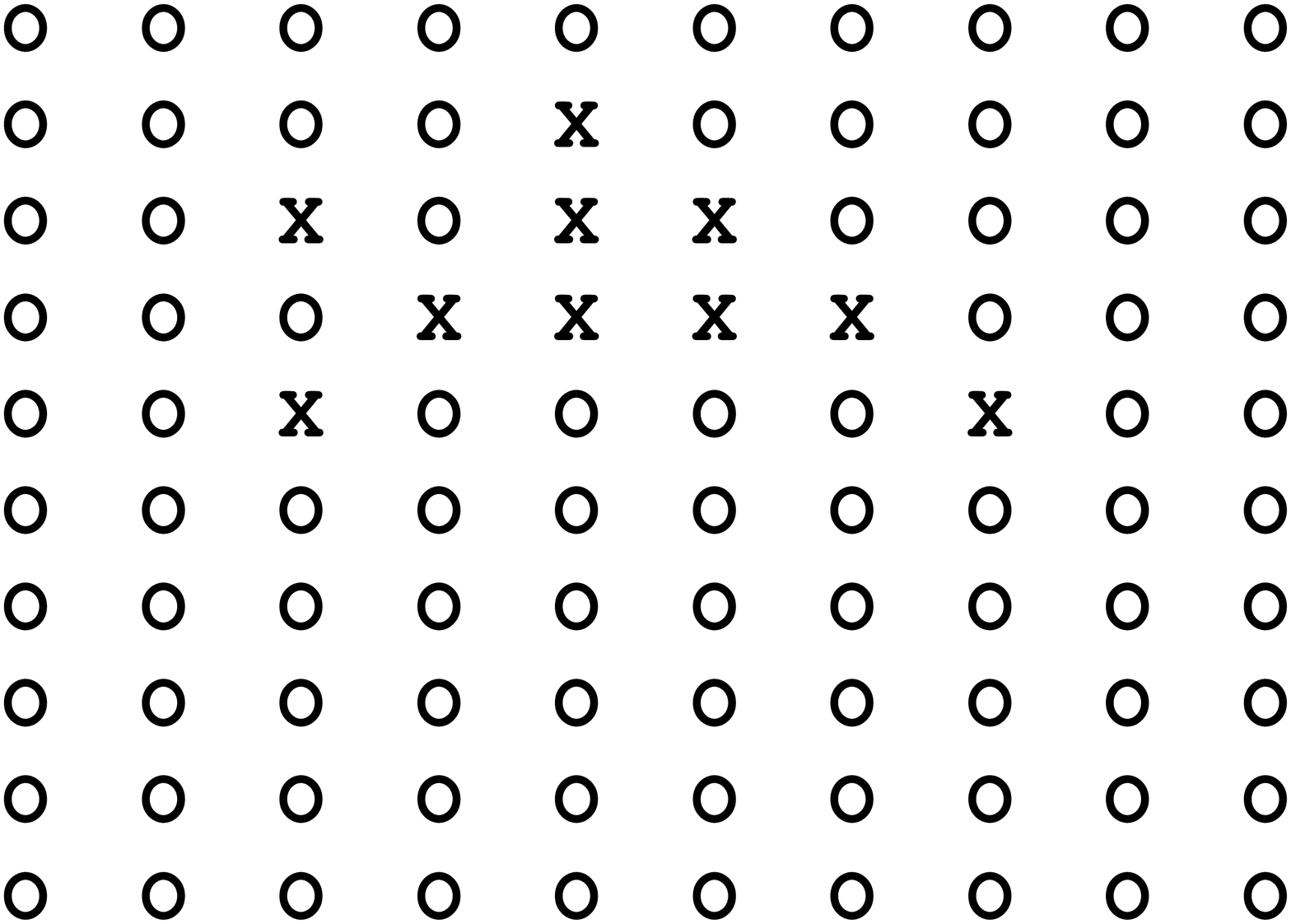}\label{10SO}}
\hspace{0.1\textwidth}
\subfigure[Random]{\includegraphics[width=0.4\textwidth]{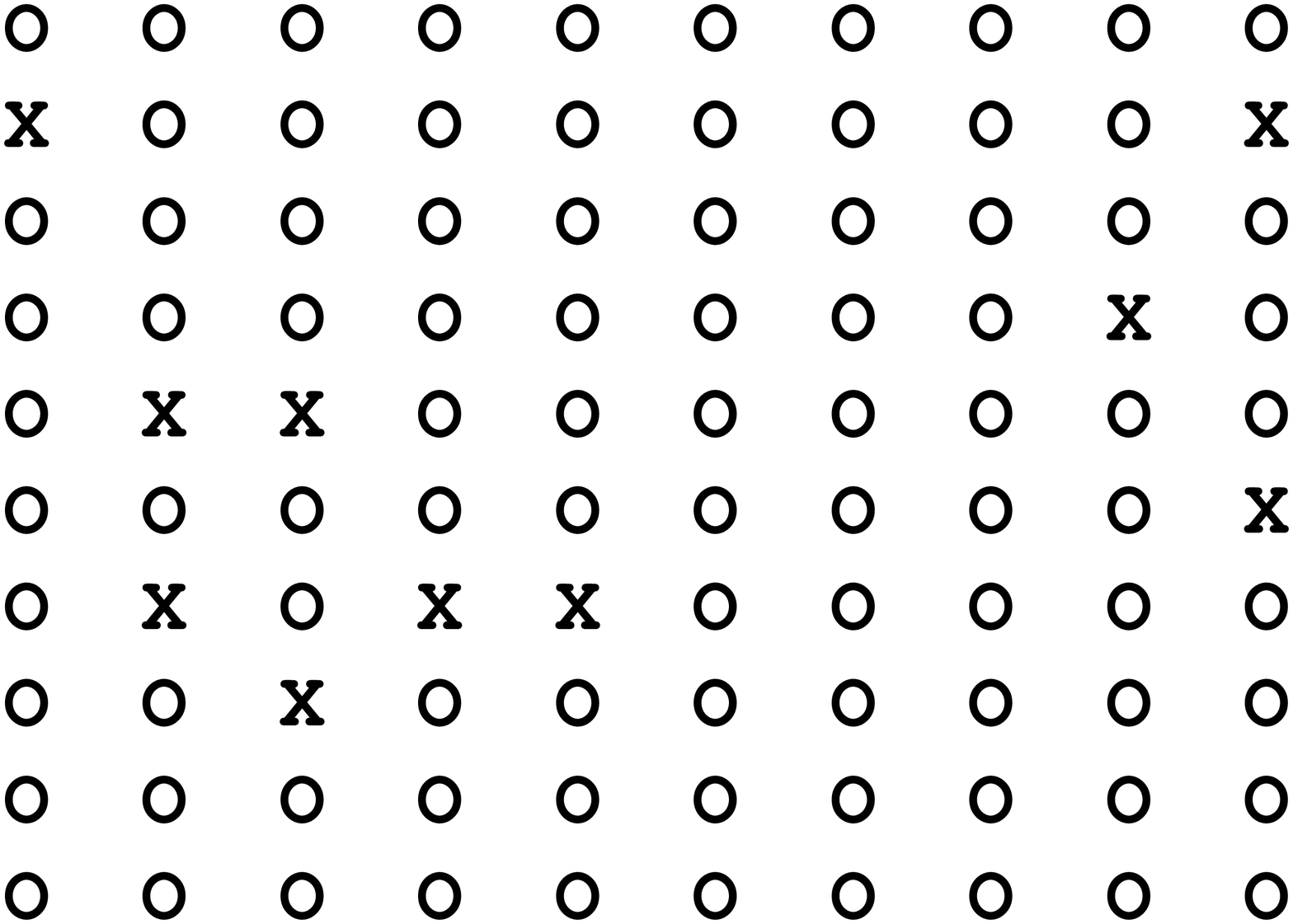}\label{10SR}}
\caption{\label{10Sites}(a) Realization of ten occupied sites in a
$10\times10$ lattice for $d=1$ (strongly correlated case); (b) the 
counterpart of a randomly populated lattice at the same density. The 
occupied sites are marked by ``X'', the empty sites by ``O''.}
\end{center}
\end{figure}

In a square lattice of linear dimension $L$, $d=L-1$ 
corresponds exactly to the uncorrelated case, whereas $d=1$ corresponds to  
the most compact population. For $d=1$, a snapshot of a realization of ten 
occupied sites on a square lattice of $L=10$ is shown in Fig.~\ref{10SO}, 
which displays obvious spatial correlations. For comparison, a randomly 
populated realization for the same density is also shown in 
Fig.~\ref{10SR}. 

Although the percolation process contains a length $d$, the correlations between site-occupation probabilities can in principle be long-ranged.  We define site-occupation variables $\theta_i$ at the sites $\{i\}$ of a square lattice.  The $\theta_i$ take on the values 1 and 0 corresponding to 
occupied and vacant sites, respectively.  Averaging over cluster realizations gives the site-occupation probability $p=\langle\theta_i\rangle$, which is the same on all sites in the growth process considered here.   To assess correlations, it is natural to compute the site-occupation correlation function:
\begin{equation}
g_\theta(\vec{r}_i-\vec{r}_j)\equiv\langle \theta_i \theta_j \rangle- \langle \theta_i\rangle\langle\theta_j\rangle
\label{corr}
\end{equation}
From the homogeneity of the percolation process, it follows that the strength of correlations between two sites $i$ and $j$ depends only on the displacement $\vec{r}_{ij}=\vec{r}_i-\vec{r}_j$.  

Now, for a finite-size system with $N$ sites, it is interesting to consider the case of site-occupation probability 
$p_N=\alpha/N$, where, the positive integer $\alpha\ll N$ does not depend on the system size $N$.  Then, the correlation function  $g_{\theta}(\vec{r}_{ij})\rightarrow -p_N^2=-\alpha^2/N^2$ tends to a non-zero constant for sufficiently large displacements $r_{ij}\gg \alpha d$.  Physically, for such a 
small site-occupation probability, only a few sites $\alpha=Np_N$ are occupied before the growth process stops.  However, if sites $i$ and $j$ are sufficiently far apart such that $r_{ij}\gg \alpha d$, then at least $r_{ij}/d\gg \alpha$ steps of the growth process are required for both sites to be occupied. 

In any realization, then, it is impossible for both sites to be occupied, once the sites are sufficiently far apart.  As a result, the product of site-occupation variables vanishes, $\theta_i\theta_j=0$, and the correlation function tends to a negative, non-zero constant $g_{\theta}(r_{ij}\gg\alpha d)=-p_N^2$.  This shows that, at least for finite systems, correlations between the site-occupation probabilities can extend over a long range, on the order of the system size.  Turning to the limit $N\rightarrow\infty$, we expect that the correlation function $g_{\theta}(\vec{r}_{ij})\rightarrow 0$ will decay to zero as $r_{ij}\rightarrow\infty$ for any $p$ and $d$.  In particular in the vicinity of percolation threshold, we anticipate that the correlations decay algebraically $g_{\theta}\sim 1/r^a$.

\section{Results}
\begin{figure}[h]
\begin{center}
\subfigure[]{\includegraphics[width=0.4\textwidth]{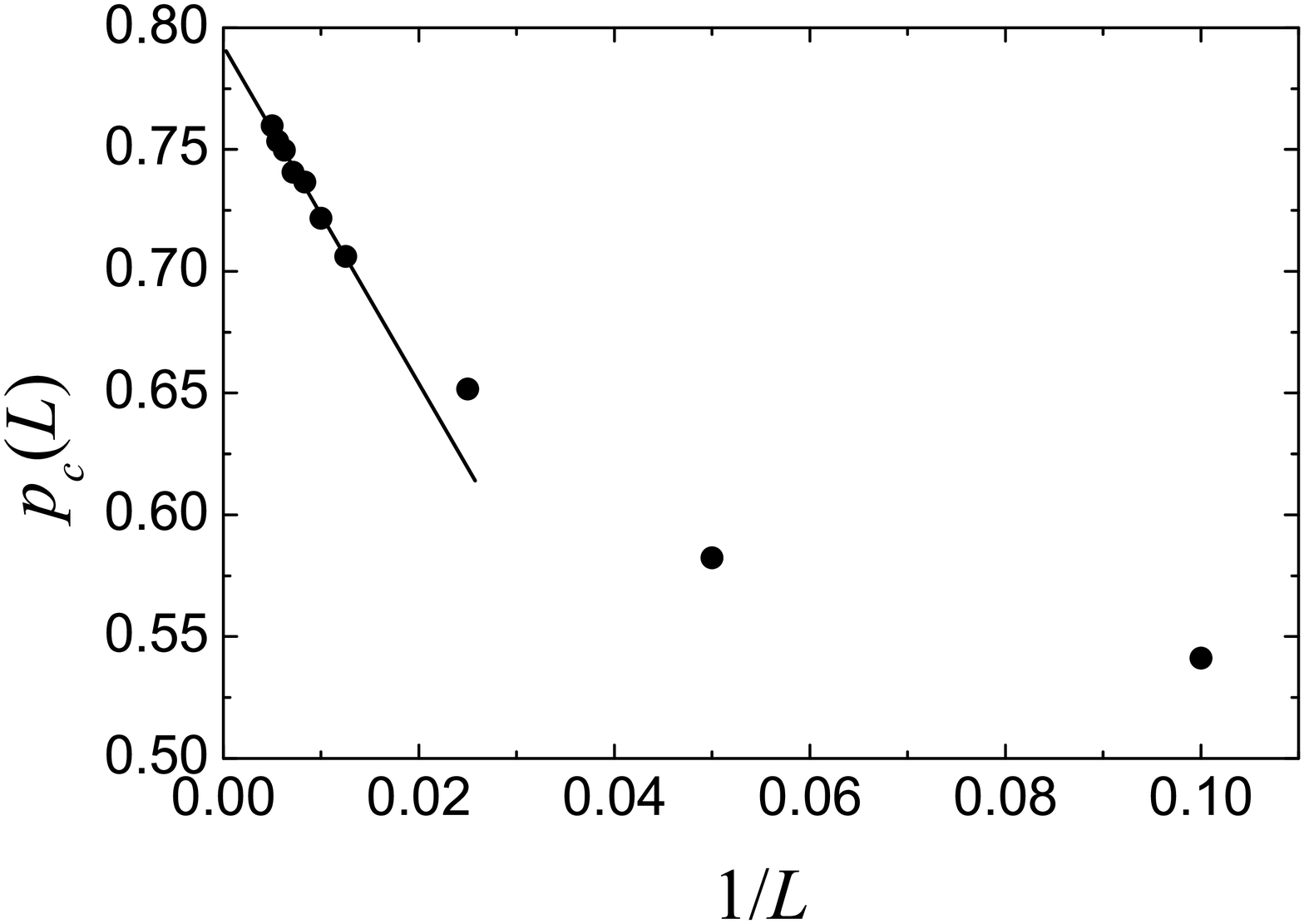}\label{pc_d1.eps}}
\hspace{0.1\textwidth}
\subfigure[]{\includegraphics[width=0.4\textwidth]{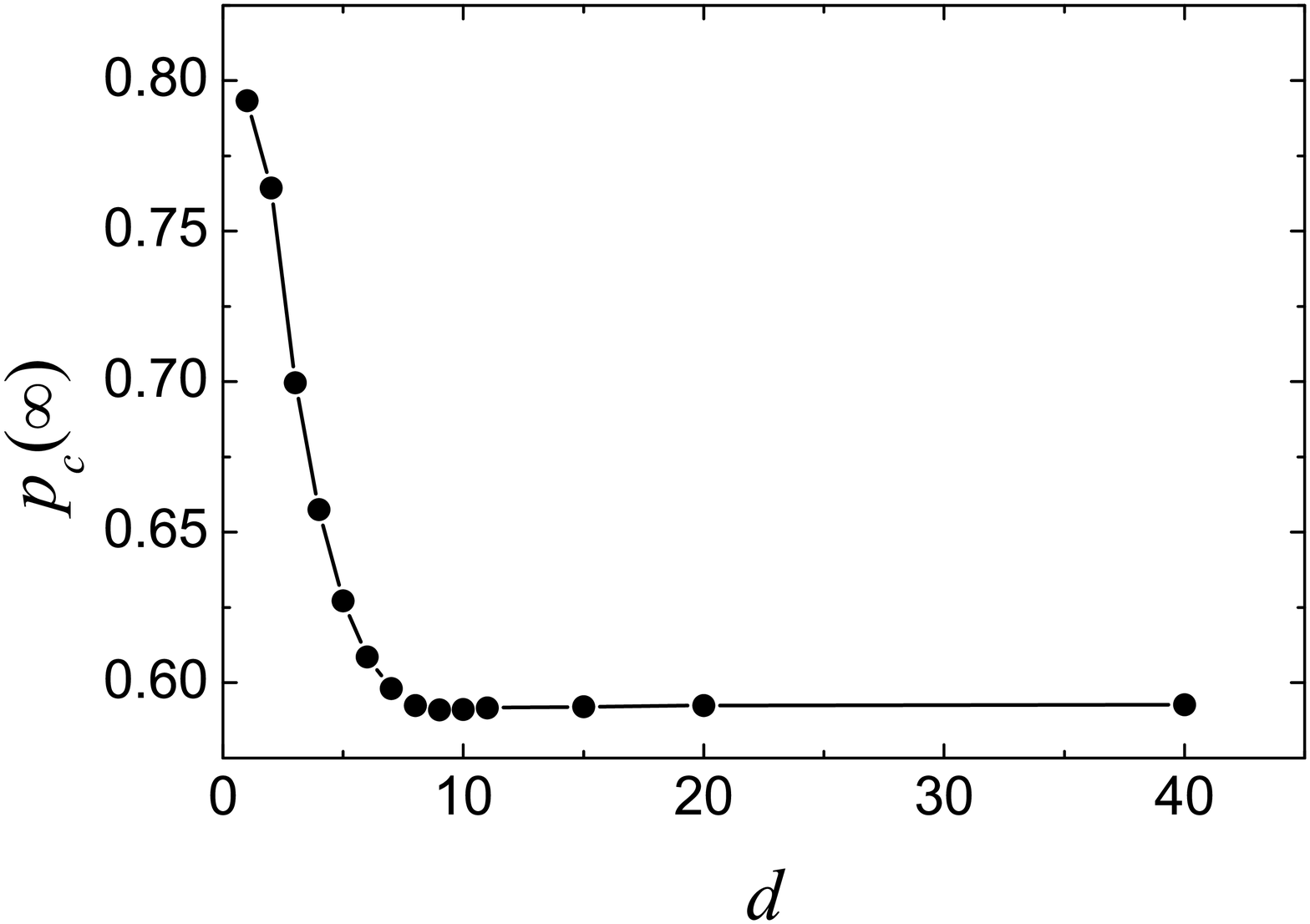}\label{pc_d.eps}}
\caption{\label{}
(a) The finite lattice percolation threshold $p_c(L)$ varies with the 
inverse of the linear dimension $1/L^{1/\nu}$ for $d=1$, with $\nu=1.0$. Linear extrapolation of 
the values of $p_c(L)$ at large $L$ yields $p_c(\infty)=0.793\pm0.002$ at 
$L=\infty$.
(b) Dependence of the percolation threshold $p_c(\infty)$ on the 
correlation parameter $d$. With increasing $d$, $p_c$ decreases quickly at 
first, reaches a minimum at about $d=9$, then grows slowly.}
\end{center}
\end{figure}

The results for percolation threshold (Fig.~\ref{pc_d1.eps}) were obtained by averaging over 3000 realizations for the largest $L=200$ system sizes\cite{GTC2006}.  For a given realization, we find the occupation fraction $p_c(L)$ when a spanning cluster first appears.  Averaging over different realizations, we arrive at a finite-size estimate that we anticipate has the following scaling behavior as $L\rightarrow\infty$: $p_c(L)=p_c(\infty)-bL^{-1/\nu}$, where, $\nu$ is the percolation length exponent\cite{NZ2001,GTC2006}.  For uncorrelated percolation, one has the exact result $\nu=\frac{4}{3}$ allowing a simple extrapolation to infinite system size.  For the correlated case, we extract the values of $\nu$ and $p_c(\infty)$ simultaneously.  

We begin by assuming $\nu=1$ (See Fig.~\ref{pc_d1.eps}) and find a preliminary estimate of $p_c(\infty)\approx 0.793\pm0.002$, for the $d=1$ data shown in the figure, using a linear least-squares fit to the finite-size scaling form.  To check for self-consistency, we take $p_c(\infty)$ in the equivalent finite-size scaling ansatz $\log( p_c(\infty)-p_c(L)) =-\frac{1}{\nu} \log L + \log b$, and use linear least-squares fitting to extract $\nu$.  The results for $p_c(\infty)$ for various $d$ shown in Fig~2(b) were obtained using $\nu=1$.  

The value of the percolation threshold for $d=1$, $p_c(\infty)=0.793\pm0.002$, is significantly larger than the uncorrelated value 
$p_c(\infty)=0.5927$\cite{NZ2001}. As $d$
increases, $p_c(\infty)$ decreases quickly, and reaches a minimum around
$d=9$, i.e. $p_c(\infty)=0.5910$. 

It is natural that at small $d$ a larger occupation fraction is required for percolation to occur.  Percolation implies that distant sites are not only simultaneously occupied, but also joined in the same cluster.  Reducing $d$, we find that distant sites have stronger anti-correlation of their occupation probabilities: they are less likely to be simultaneously occupied.   

It is noteworthy that the percolation threshold $p_c(\infty)$ in Fig. 2(b) gently increases with $d$ beyond $d=9$.  For large $d$, an increasing number of the randomly added sites are not part of the spanning cluster, and therefore percolation requires a larger population density.  Numerically we find that for $d=20$, the percolation threshold is slightly smaller $p_c(\infty)=0.5924$ than the threshold at  $d=40$, $p_c(\infty)\approx0.5927$.  The latter is indistinguishable from the uncorrelated result, at this level of precision.  It would be interesting to consider other correlated percolation processes for which this dip in $p_c(\infty)$ is more pronounced.   

Remarkably, the extracted values of $p_c(\infty)$ depend very weakly on $\nu$, allowing for a precise estimate despite the large uncertainty in $\nu$: We find self-consistent fits using $\nu=1.0\pm0.3$ for all $d$.  Further work is necessary to provide a reliable estimate of the correlated percolation length exponent $\nu$.  Weintraub \cite{W1984} has suggested the result $\nu_{long}=2/a$, where, the exponent $a$ governs the decay of correlations $g_{\theta}\sim 1/r^a$ at large distance.  It would be interesting to compare more precise estimates of $\nu$ to this prediction, assuming that the decay of the correlations takes this form.

\begin{figure}[h]
\begin{center}
\subfigure[]{\includegraphics[width=0.4\textwidth]{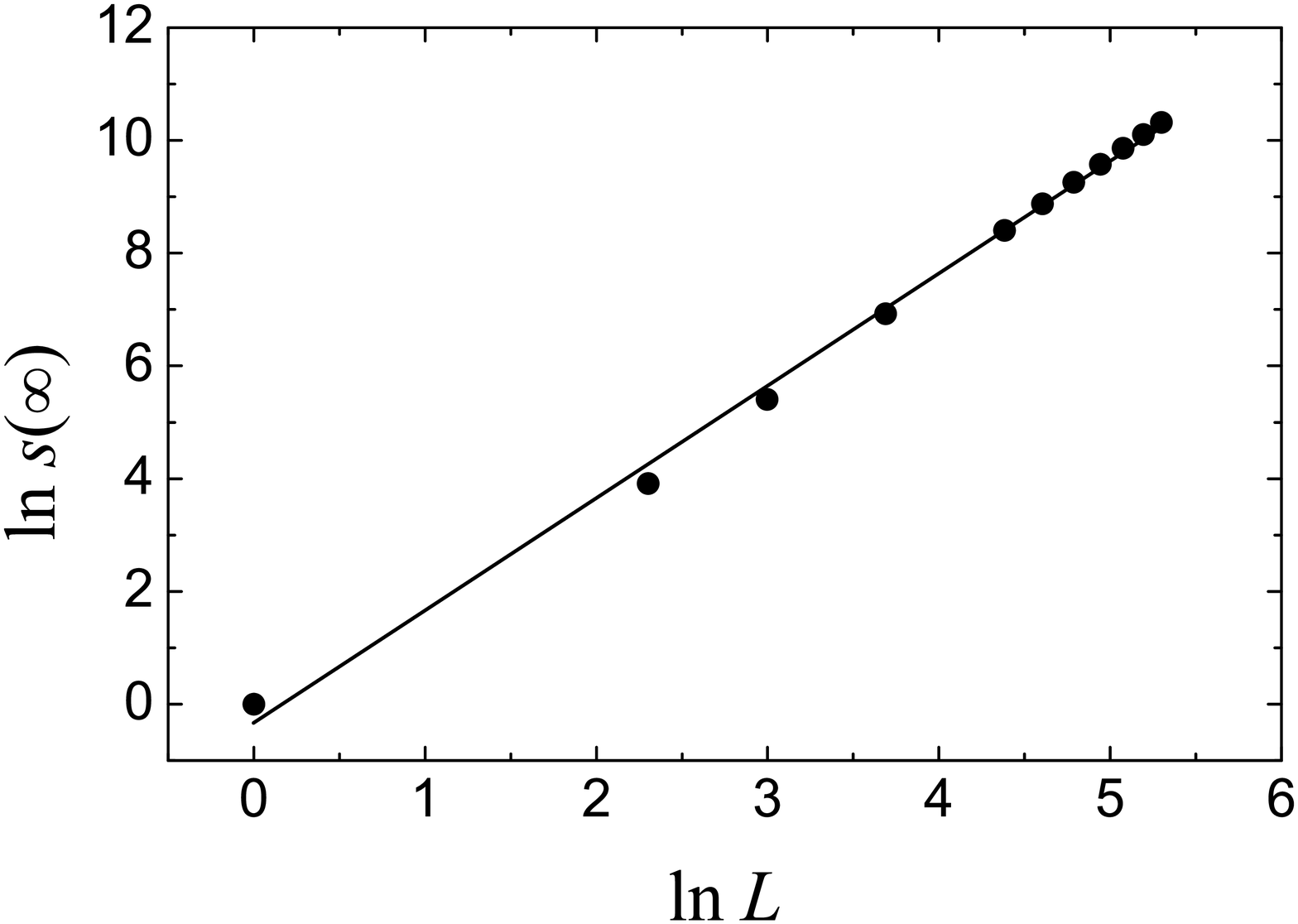}\label{d1_D.eps}}
\hspace{0.1\textwidth}
\subfigure[]{\includegraphics[width=0.4\textwidth]{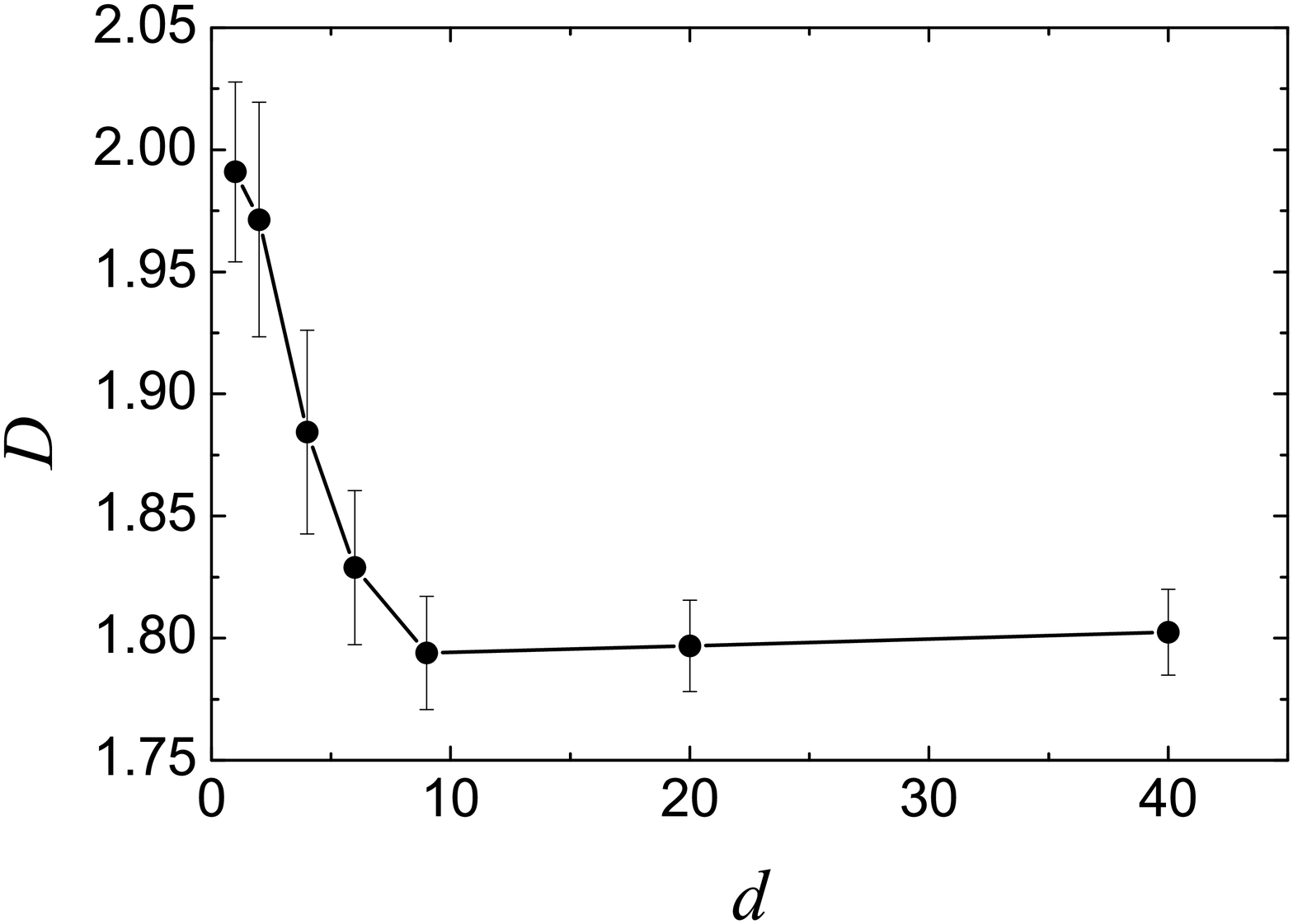}\label{D(d).eps}}
\caption{\label{}(a) The logarithm of the spanning cluster size at the 
correlated percolation threshold for $d=1$, plotted as a function of the 
logarithm of the linear dimension. A linear fit of this plot gives 
the slope $1.99\pm0.04$. 
(b) The fractal dimension $D$, as a function of 
the correlation parameter $d$. The fitting errors are also given.}
\end{center}
\end{figure}

To gain a better understanding of the effect of correlations near threshold, we turn to the fractal
dimension $D_f$, governing the growth of the number of sites in the spanning cluster $s(\infty)\propto{L}^{D_f}$ \cite{SA1992}. In Fig.~\ref{d1_D.eps}, we plot the size of the spanning cluster on a log-log scale as a function of system size at fixed correlation
$d=1$.  The slope of the line
gives $D_f=1.99\pm0.04$ for $d=1$, indistinguishable from the Gaussian fixed point result $D_f=2$.  Similarly, we obtain
the fractal dimensions corresponding to other correlation strengths.  The dependence of the fractal dimension on the correlation parameter
$d$ is illustrated in Fig.~\ref{D(d).eps}. This plot shows a
dependence similar to that of correlated percolation thresholds in Fig.~2(b).
Increasing $d$ from 1 to 9, the fractal dimension of the spanning cluster decreases, until a shallow minimum around $d=9$, after which the fractal dimension increases towards the uncorrelated value.
We note however a quantitative difference: the fractal dimension $D_f=1.80\pm0.02$ for the largest $d=40$ data shown is still far less than the exactly known value for
uncorrelated percolation $91/48\approx 1.896$. This is to 
be contrasted with the numerically obtained percolation threshold at 
$d=40$, which coincides with that of uncorrelated percolation, within the precision of our numerical calculations.

\section{Discussion}
We interpret the results for the fractal dimensions as an indication for a crossover, as function of $d$, between different renormalization-group fixed points.  The fractal dimension $D_f=D/2+1-\eta/2$, can be expressed in terms of the anomalous dimension $\eta$ governing the probability that two sites belong to the same cluster.  Close to the upper critical dimension $D=6$, three fixed points dominate the phase diagram for long-range correlated percolation\cite{W1984}.  For the Gaussian fixed point, the anomalous exponent $\eta_{gauss}=0$ vanishes.  For the pure fixed point (at which correlations between couplings in the Potts model are irrelevant) and the long-range fixed point, $\eta$ is non-zero.  In contrast to the pure and Gaussian fixed points, the long-range value depends on $a$: $\eta_{long}=(D-2-a)/11$ for $a\approx 4$.  Above the upper critical dimension $D>6$, the Gaussian fixed point governs for $a\nu_{gauss}-2>0$.  Below the upper critical dimension $D<6$, the pure fixed point governs once $a\nu_{pure}-2>0$.  The long-range fixed point governs when neither the pure nor the Gaussian do.  Assuming these results extend qualitatively to the lower critical dimension $D=2$, one may interpret our numerical results as follows.
  
For $d\ll 10$, we observe $D_f=2$, corresponding to a vanishing anomalous dimension.  This is naturally interpreted as a Gaussian fixed point.  In the limit $d=\infty$, our growth process is identical to uncorrelated percolation and is governed by the pure fixed point with $D_f=\frac{91}{48}$.  In between these two extremes, for large but finite $d$, our numerical results are inconsistent with both the pure and the Gaussian fixed points, and are likely to be governed by a long-ranged fixed point with fractal dimension that depends on $d$ through the exponent $a$.  Within this picture, the non-monotonic dependence of $D_f$ on $d$ results from the presence this new, long-ranged fixed point at intermediate $d$.  

In conclusion, we have generalized the Monte Carlo method by Newman
and Ziff using a simple correlated percolation model. We have then
calculated percolation thresholds, 
spanning cluster sizes and fractal dimensions for different correlation 
ranges.  The observed non-monotonic dependence of the percolation threshold on the 
correlation range can be understood in terms of two competing 
mechanisms: compactification of the spanning cluster as $d\rightarrow 1$ 
and increasing abundance of particles that are not part of the 
spanning cluster as $d\rightarrow \infty$.   The dependence of the fractal dimension $D_f$ on $d$ is less trivial.  We have presented a scenario involving a cross-over, as a function of $d$, between three renormalization group fixed points.  It would be interesting to 
apply this analysis to physical systems with tunable spatial correlations in the site-occupation probabilities, 
using it to control and optimize their critical properties.

\section{Acknowledgements}
The authors thank Letian Ding and Silvano Garnerone for helpful 
discussions. H.Y. thanks Farhad Arbabzadah
for sharing his experience in programming and knowledge in percolation.  This work was supported by DOE grant DE-FG02-06ER46319 and the China Scholarship Council (H.Y.).

\end{document}